# DuAL-Net: A Hybrid Framework for Alzheimer's Disease Prediction from Whole-Genome Sequencing via Local SNP Windows and Global Annotations

Eun Hye Lee[a], Taeho Jo[a*]

[a] Indiana Alzheimer Disease Research Center and Center for Neuroimaging, Department of Radiology and Imaging Sciences, Indiana University School of Medicine, Indianapolis, IN 46202, USA

## Abstract

Alzheimer's disease (AD) dementia is the most common form of dementia. With the emergence of disease-modifying therapies for AD such as anti-amyloid monoclonal antibodies, the ability to predict disease risk before symptom onset has become increasingly important. Whole genome sequencing (WGS) data is a promising data form for early AD prediction, despite several analytical challenges. In this study, we introduce DuAL Net, a hybrid deep learning framework designed to predict AD dementia using WGS data.

DuAL-Net integrates two components: local probability modeling, which segments the genome into non-overlapping windows, and global annotation-based modeling, which annotates each SNP and reorganizes the WGS input to capture long range functional relationships. Both components employ out of fold stacking with TabNet and Random Forest classifiers. The final prediction is generated by combining local and global probabilities using an optimized weighting parameter alpha.

We applied DuAL-Net to WGS data from 1,050 individuals (443 cognitively normal and 607 with AD dementia), using five-fold cross validation for training and evaluation. On average across the 100, 500, and 1000 SNP subset sizes evaluated, DuAL-Net achieved an Area Under the Curve (AUC) of 0.671 using top-ranked SNPs prioritized by the model, representing 35.0% and 20.3% higher predictive performance compared to the average AUCs of bottom-ranked and randomly selected SNPs, respectively. Assessment of model discriminative ability via ROC analysis across different SNP subset sizes consistently demonstrated a strong positive correlation between the SNPs' prioritization rank and their predictive power. The model identified SNPs with known associations to AD as top contributors to prediction, alongside potentially novel variants also ranked highly by the model.

In conclusion, DuAL Net presented a promising framework for AD prediction that improved predictive accuracy and enhanced biological interpretability. The framework and its web-based implementation offer an accessible platform for broader research applications.

**Introduction**

Alzheimer's disease (AD) dementia, the most common type of dementia, is a progressive neurodegenerative disorder characterized by the accumulation of amyloid plaques and neurofibrillary tangles in the brain.[1] With the emergence of anti-amyloid monoclonal antibody therapy for AD, early prediction of individuals at risk, especially before the onset of clinical symptoms, has become increasingly important, as these therapies are suggested to be more effective when initiated in the earlier stages of disease progression.[2, 3] However, there are currently no effective screening methods, particularly before clinical and pathological changes manifest.

In this context, genomic data presents a promising avenue for the early prediction of AD. Germline genomic data remains unchanged throughout an individual's lifetime, enabling risk prediction from birth if accurate predictive models are developed. Additionally, twin studies estimate the genetic heritability of AD to be 50% to 70%,[4, 5] highlighting a strong genetic component and the potential of genomic data as a predictive tool.

Whole genome sequencing (WGS) data offers more comprehensive information compared to other types of genomic data even though its high dimensionality poses a significant challenge for analysis.[6] Moreover, the hierarchical structure of the genome presents an additional challenge, as both local interactions between nearby loci and long-range relationships between distant genomic regions can influence the endophenotype.[7] However, encoding or embedding methods that rely solely on raw WGS input often struggle to effectively capture these long-range relationships.

To address these limitations of WGS analysis, we proposed a hybrid framework that incorporates analyses of both local and long-range relationships. We reduced input dimensionality by dividing the WGS data into non-overlapping local windows. In parallel, we annotated each SNP and reorganize the input based on these annotations, allowing the model to capture long-range relationships. This strategy possibly reflects the genome's hierarchical structure and enables the model to improve disease prediction performance for complex disease such as AD.

Currently, most of the genome-based models yield only modest performance for predicting AD.[8, 9] To address this challenge, our novel framework applied ensemble framework that combined two fundamentally different models. Random Forest (RF), a decision tree–based method known for its robustness across various genomic applications,[10-12] and TabNet, a deep learning model optimized for tabular data with built-in interpretability.[13, 14] By integrating RF and TabNet through a stacking ensemble, the model leverages their complementary strengths, thereby enhancing overall performance.

In this study, we aim to (1) develop a novel hybrid framework optimized for WGS data analysis, (2) validate its potential for the prediction of AD dementia, and (3) identify novel SNPs that contribute significantly to AD dementia prediction within the context of the input WGS data. By introducing this hybrid framework, we provide a practical approach for leveraging WGS data in disease prediction, with the goal of improving the scalability and applicability of genomic models for the early identification of individuals at risk for AD dementia.

**Methods**

**Study participants**

This study utilized whole genome sequencing (WGS) data from total of 1,566 participants drawn from two sources: the Alzheimer's Disease Neuroimaging Initiative (ADNI, n=809)[15] and theADNI-WGS-2 with Alzheimer's Disease Sequencing Project (ADSP) Follow-Up Study (ADSP-FUS1-ADNI-WGS-2, n=757).[16,17] ADNI, launched in 2003 and extended through its phases (ADNI-1, ADNI-GO, ADNI-2, ADNI-3), aims to discover biomarkers and monitor AD progression. This includes utilizing serial imaging studies, various biological markers, clinical and neuropsychological evaluations, as well as genomic information including WGS. The ADSP Follow-Up Study builds on the ADNI framework by extending longitudinal follow-up intervals and adding whole-genome sequencing data for participants not previously sequenced, thus offering broader genomic coverage for AD-related investigations. From the combined pool, individuals diagnosed with MCI were excluded, yielding a final set of 443 cognitively normal (CN) participants and 607 with clinically confirmed AD dementia. Demographic information, *APOE* and whole genome genotyping data, and clinical information were obtained from the ADNI data repository (www.loni.usc.edu/ADNI/). All participants provided written informed consent, and the study protocol was approved by the institutional Review Board at each data acquisition site. We selected CN and AD participants and their demographic information included in this study is shown in Table 1.

*Table 1. Demographic characteristics and APOE genotype distribution for participants*

|  | CN (N=443) | AD dementia (N=607) |
|---|---|---|
| **Age** | 72.6 (6.3) | 74.0 (7.3) |
| **Sex** | | |
| Male | 235 (53.0) | 255 (42.0) |
| Female | 208 (47.0) | 352 (58.0) |
| *APOE* **genotype** | | |
| ε2/ε2 | 2 (0.5) | 1 (0.2) |
| ε2/ε3 | 57 (12.9) | 19 (3.1) |
| ε2/ε4 | 7 (1.6) | 14 (2.3) |
| ε3/ε3 | 257 (58.0) | 194 (32.0) |
| ε3/ε4 | 110 (24.8) | 278 (45.8) |
| ε4/ε4 | 10 (2.3) | 101 (16.6) |

*Values are presented as mean (SD) or number (%).*

*Abbreviations: AD, Alzheimer's disease; CN, cognitive normal; SD, standard deviation*

**Whole-genome sequencing data**

This study leveraged newly integrated WGS data from both ADNI and the ADSP-FUS1-ADNI-WGS-2 cohort, thus extending the follow-up period and augmenting the genetic dataset for AD research. Sequencing was primarily conducted on Illumina platforms (HiSeq2000, HiSeq X, or NovaSeq) with read lengths of either 100 bp or 150 bp. Library preparation was performed using standard Illumina reagents, and the resulting paired-end reads were aligned to the GRCh38 (hg38) human reference genome with BWA-MEM.[18] Picard tools were used to mark PCR duplicates and collect alignment metrics, followed by local realignment around insertions or deletions and base quality score recalibration with the Genome Analysis Toolkit (GATK).[19] Joint variant calling across all samples proceeded via the GATK HaplotypeCaller, adhering to best-practice guidelines, and the Genome Center for Alzheimer's Disease (GCAD) implemented its Variant Calling Pipeline (VCPA) v1.0 to standardize quality control (QC) measures.[20, 21] Initial QC steps included verifying SNV concordance, identifying sex mismatches, detecting contamination, and evaluating relatedness (Pihat > 0.4). Additionally, sample-level QC excluded individuals with sex inconsistencies, call rates below 95%, or duplicated genetic profiles. At the SNP level, variants with a call rate below 95%, Hardy–Weinberg equilibrium p-values $< 1 \times 10^{-6}$, or minor allele frequency (MAF) < 1% were removed. Further filters discarded genotypes with low genotype quality (GQ < 20) or insufficient read depth (DP < 10), and set to missing those variants exhibiting missing call rates > 20%. Finally, any variants presenting as monomorphic or multi-allelic were also excluded.

**Imputation**

In our previous work, we systematically evaluated multiple imputation algorithms, including k-nearest neighbors (k=1, 5, 10), a Generative Adversarial Network (GAN) Imputer, an Iterative Imputer, MissForest, and a Simple Imputer.[22] This evaluation involved artificially inflating the proportion of missing genotypes and subsequently comparing reconstructed data using accuracy, Root Mean Squared Error (RMSE), R-squared ($R^2$), Mean Absolute Error (MAE), and Normalized RMSE (NRMSE). In the current study, we adopted the same pipeline to ensure methodological consistency with our previous approach and to enable a direct comparison of imputation quality.

After imputation, a $1 \times 10^6$ base pair region of WGS data encompassing the *APOE* gene on chromosome 19 was extracted for analysis. This region included 14,011 SNPs.

**Genomic information of SNPs**

All retained SNPs were annotated using Ensembl genome annotation resources (Ensembl release 108, GRCh38). Figure 1 shows the annotation steps, which included determining genomic context (exonic, intronic, untranslated regions), regulatory elements (promoters, enhancers), predicted variant consequences, epigenetic features, and clinical significance categories. We leveraged the pyensembl library and Ensembl's RESTful APIs to systematically retrieve gene-, transcript- and variant-level annotations.

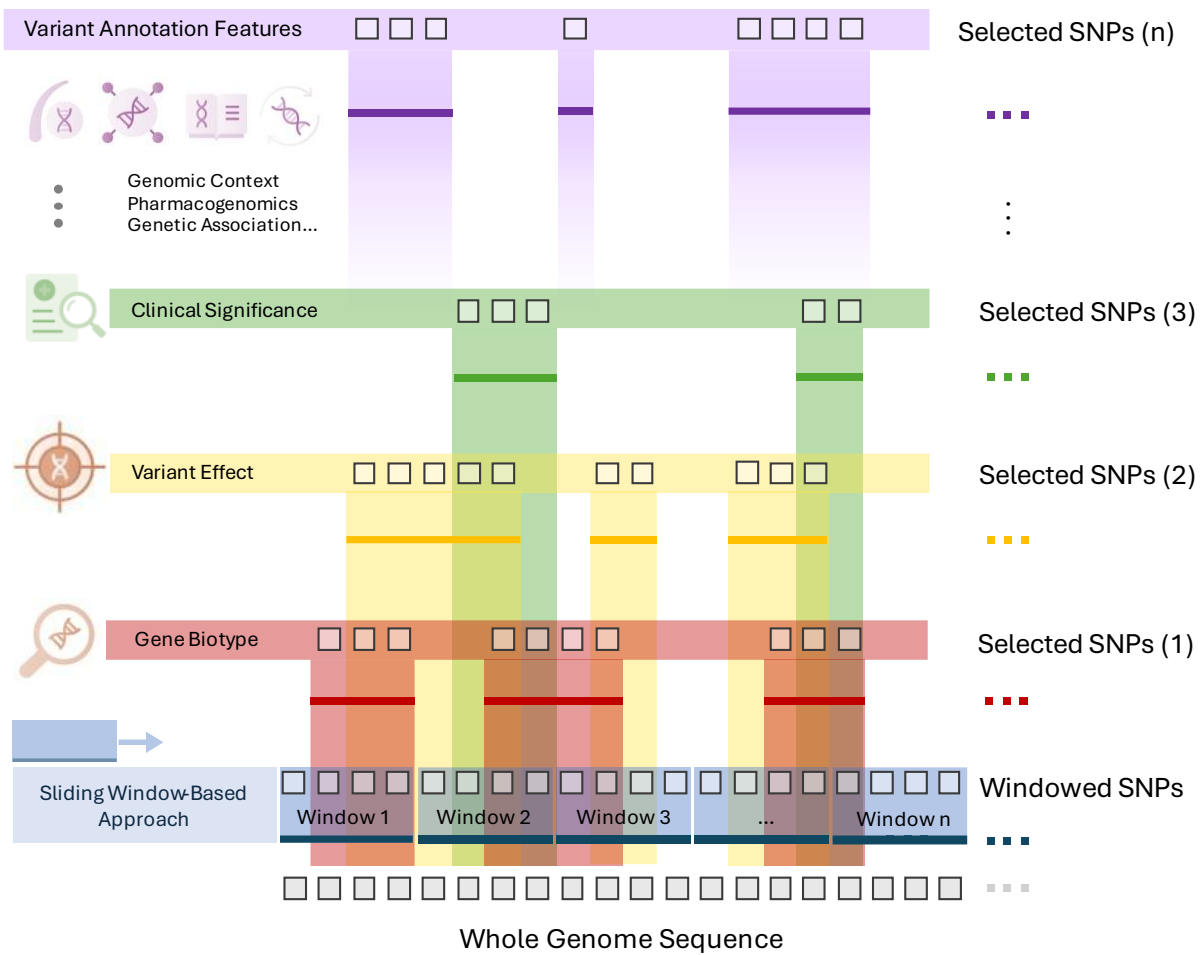

**Figure 1 Annotation of genomic information to each SNP and grouping of SNPs by genomic windows or annotation categories**

*After imputation, annotation was performed on the selected WGS data within a $1 \times 10^6$ base pair region surrounding APOE region using Ensembl's RESTful APIs. Each SNP was divided into non-overlapping 100-base pair windows using a sliding window approach, and additionally grouped by annotation categories. The groups based on non-overlapping windows and annotation were used as inputs for a model that combines TabNet and Random Forest via out-of-fold stacking.*

*Abbreviations: SNP, single nucleotide polymorphism*

**Model Architecture**

Our model architecture, which we named DuAL-Net (Dual Approach Local-global Network for Alzheimer's Risk Assessment), integrated TabNet[14] and RF to generate out-of-fold (OOF) predictions, which were subsequently used for meta-model training. The DuAL-Net framework was designed to leverage both local genomic context and global functional annotations to improve AD prediction accuracy. The complete implementation is available as open-source software on GitHub (https://github.com/taehojo/DuAL-Net) and accessible through a web server (https://www.jolab.ai/dualnet).

**Local Probability Modeling with Out-of-Fold Stacking**

Local SNP interactions within genomic regions were captured through an out-of-fold stacking ensemble strategy applied to sliding windows of SNPs. After QC, individual SNP data was fragmented into non-overlapping 100-base pair sized windows. The SNPs within each window were used as features for the classification model. Two different classifiers were trained to predict AD dementia (versus CN) using the SNPs within each window: a TabNet model, which employs a sequential attention mechanism designed for tabular data that selectively focuses on informative features, and a RF classifier, which is an ensemble of decision trees. To compute local accuracy, we first generated OOF predictions using both TabNet and RF within a 5-fold cross-validation (CV) framework. The 1,050 samples were randomly divided into five folds for cross-validation. In each iteration, four folds were used for training the TabNet and RF models, with the remaining fold held out for validation. The OOF accuracy estimates from both models were then used as features to train a logistic regression meta-model, producing a final local accuracy score, representing the likelihood of AD diagnosis.

**Global Annotation-Based Modeling**

Annotation data from the previous 'Genomic information of SNPs' section were converted into binary indicator variables. This encoding facilitated the selection of groups of SNPs sharing specific annotations. Specifically, we utilized the following annotation categories: genomic context (region none, indicating SNPs not located in exonic, intronic, or untranslated regions [UTRs]); gene biotype (TEC, lncRNA, protein coding, transcribed unprocessed pseudogene, unknown); most severe consequences (3' UTR variant, 5' UTR variant, transcription factor binding site variant, unknown, intergenic variant, intron variant, missense variant, regulatory region variant, splice donor region variant, splice polypyrimidine tract variant, splice region variant, stop gained, synonymous variant); and clinical significance (pathogenic, likely pathogenic, uncertain significance, drug response, other, risk factor, association, protective, established risk allele). For each annotation category, such as SNPs in protein-coding regions, SNPs with a missense consequence, or clinically pathogenic SNPs, a classifier was trained and evaluated using only those SNPs. The same TabNet and RF stacking procedure described above was applied to these

SNP subsets in a five-fold cross-validation setting. This method produced a classification accuracy and receiver operating characteristic area under the curve (ROC AUC) for each annotation-defined group, reflecting the predictive information captured by the SNPs sharing that annotation. Each SNP then received a global accuracy score calculated by averaging the classification accuracies of all annotation groups in which it appeared.

**Combining Local and Global Accuracy**

The local and global scores were subsequently merged to produce a combined SNP prioritization metric. To obtain a final probability estimate, we combined local and global probabilities using a weighting parameter α in the range [0,1], which determines the relative contribution of each score. The final combined score is computed as follows:

Combined score = α × local accuracy score + (1−α) × global accuracy score

An α of 1.0 used only the local model score, whereas an α of 0.0 used only the global score, and intermediate values provided a linear blend of the two.

To determine the optimal α, the analysis considered values in increments of 0.1. For each candidate α, all SNPs were re-ranked according to their combined score, and the top 100 SNPs under that weighting were selected. A stacked classifier, restricted to these 100 SNPs, was then trained and evaluated via cross-validation to determine its ability to distinguish AD cases from controls. The α that produced the highest average AUC across folds was selected as the best trade-off between local and global information.

**Model Evaluation and SNP Prioritization**

After selecting the optimal α, all SNPs were assigned final combined scores and ranked from most to least likely to be associated with AD. To evaluate the methodology, we created multiple SNP panels of varying sizes (100, 500, and 1000) from the top-ranked SNPs. For comparison, we also created corresponding panels of the same sizes using the bottom-ranked SNPs and randomly selected SNPs. Classification models were trained on these respective sets using a five-fold cross-validation framework. ROC curves were generated by aggregating out-of-fold predictions, and the AUC was calculated to quantify the predictive capacity of each SNP panel. Performance metrics were derived from cross-validated predictions to maintain methodological rigor.

**Results**

**Local SNP Window Analysis**

We evaluated our DuAL-Net framework on 1,050 samples using an out-of-fold stacking ensemble approach. The genome was divided into 140 contiguous windows of 100 SNPs each. The average classification accuracy across all windows was 0.582. Among the 140 windows evaluated, the lowest accuracy was 0.543 and the highest reached 0.676. A total of 89 windows produced accuracies between 0.57 and 0.58, while 22 windows achieved accuracies of 0.60 or higher. This approach identified genomic segments with above-average accuracy that were carried forward to the integrated analysis.

**Global Annotation-Based Analysis**

We performed a global analysis using functional annotations for each SNP. Analysis of individual annotation features revealed varying predictive power. SNPs with ClinVar clinical significance of "pathogenic" achieved the highest classification accuracy at 0.686. Variants labeled as "uncertain significance," "likely pathogenic," or "drug response" yielded accuracies of 0.676, 0.676 and 0.667, respectively. SNPs located in protein-coding genes predicted AD with 0.657 accuracy. The global analysis identified several annotation features with above-average predictive power for disease status.

**Integration of Local and Global Scores (α Optimization)**

Through cross-validation, we determined that combination weight $\alpha = 0.8$ yielded optimal performance. Under this setting, the DuAL-Net, an integrated approach, achieved a cross-validated accuracy of 0.678. The global-only model ($\alpha = 0.0$) produced an accuracy of 0.605, while the local-only model ($\alpha = 1.0$) achieved 0.677.

**SNP Ranking and ROC Performance**

Using the optimized combined score ($\alpha = 0.8$), we ranked all SNPs from highest to lowest scores (Supplementary Table 1). To evaluate the discriminative power of this ranking, we conducted ROC analyses on different subset sizes (100, 500, and 1000 SNPs). For each subset size, we compared three groups: (1) top-ranked SNPs, (2) bottom-ranked SNPs, and (3) randomly selected SNPs.

For the 100-SNP subset, the top-ranked SNPs achieved an AUC of 0.697, while bottom-ranked and random SNPs yielded AUCs of 0.473 and 0.456, respectively. Across the tested subset sizes, the average AUCs were 0.671 for top-ranked, 0.497 for bottom-ranked, and 0.558 for random SNPs. Table 2 and Figure 2 show the pattern that top-ranked SNPs consistently presented higher AUC values than bottom-ranked or random variants across other subset sizes.

**Table 2. Comparison of accuracies by DuAL-Net with top-ranked, bottom-ranked and randomly selected SNP subsets**

| SNP Subset Type | AUC (N=100) | AUC (N=500) | AUC (N=1000) | Average AUC |
| --- | --- | --- | --- | --- |
| top-ranked SNPs | 0.697 | 0.660 | 0.657 | 0.671 |
| bottom-ranked SNPs | 0.473 | 0.506 | 0.513 | 0.497 |
| randomly selected SNPs | 0.456 | 0.598 | 0.621 | 0.558 |

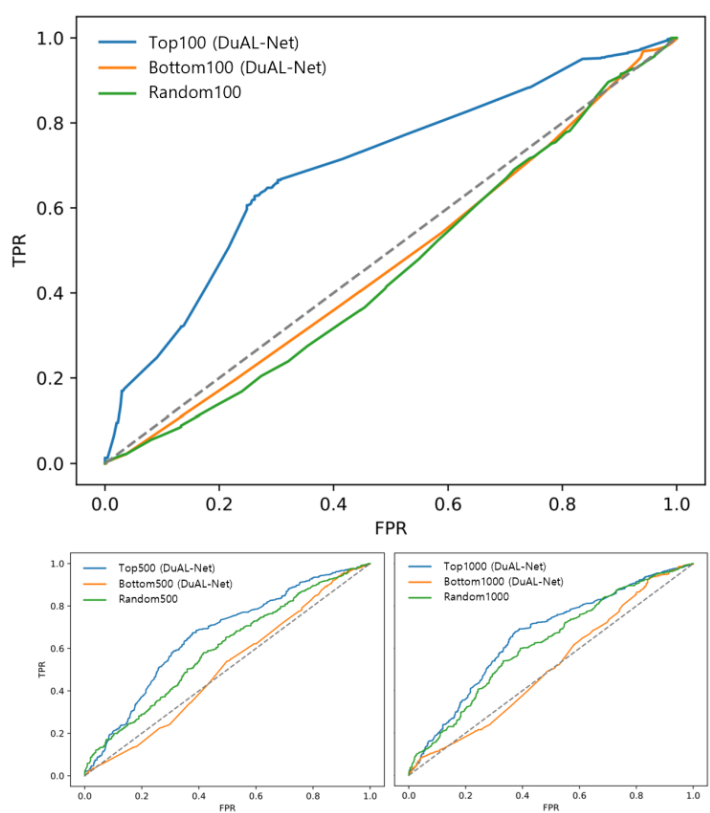

**Figure 2. ROC curves comparing top-ranked, bottom-ranked and randomly selected SNP subsets identified by DuAL-Net**

*To evaluate the discriminative power of the ranking calculated by DuAL-Net, we compared ROC curves among three groups (top-ranked SNPs, bottom-ranked SNPs, randomly selected SNPs) within different subset sizes (100, 500 and 1000 SNPs). The top-ranked SNPs presented higher AUC values than bottom-ranked or randomly selected SNPs throughout all subset sizes.*

*Abbreviations: FPR, false positive rate; ROC, receiver operating characteristic; SNP, single nucleotide polymorphism; TPR, true positive rate*

A total 72 genes were included in the 1 × 10⁶ base pair segment of WGS data surrounding the APOE gene. Among these, top 100-ranked SNPs were only from *APOE, APOC1* regions. When it extended to top 500-ranked SNPs, *TOMM40, NECTIN2* also included. Within top 100-ranked SNPs, rs429358 and rs7412 were identified, and rs34954997 and rs72654445, related to lipid metabolism, were also included. In the top 500 SNPs, an additional 23 SNPs were identified in relation to AD, dementia, or memory impairment, including rs157588, rs11668327, and rs157590. rs1064725, rs12721054, rs142042446, and rs12721052 were also observed among the top 500 SNPs, which have not been reported to be associated with AD or cognitive function in previous GWAS.

**Discussion**

This study proposed DuAL-Net as a new hybrid model based on TabNet optimized for analyzing WGS data to predict AD. By combining local SNP window analysis with global annotation-based analysis, DuAL-Net achieved more accurate AD prediction than approaches using local SNP window analysis alone. The model showed that selecting top-ranked SNPs provided higher predictive performance than random or bottom-ranked selections. However, in larger subsets, random choice captured some predictive SNPs by chance, narrowing the performance gap. This underscores DuAL-Net's role in identifying important SNPs for AD prediction. Furthermore, SNPs identified as highly important by the DuAL-Net model captured established disease-associated genomic variants, supporting the biological relevance of the model's outcomes.

The DuAL-Net demonstrate that integrating complementary sources of genomic information improves predictive performance. While the local SNP window analysis and the global annotation-based analysis each achieved moderate accuracy, the integrated model with optimized weighting (α = 0.8) attained a higher accuracy of 0.678 compared to them. This improvement indicates that combining both analyses effectively captures the multilayered structure of WGS data and provides additional information for AD prediction. By using external annotations to guide SNP selection, we created a biologically informed framework that reduces the WGS feature space and employs a hierarchical approach, enhancing both interpretability and computational efficiency. Furthermore, our framework provides a generalizable strategy for incorporating diverse multimodal data into WGS analysis. DuAL-Net features a simpler architecture that is less computationally demanding and can accommodate more than three data types, compared with cross-attention mechanisms employed in some recent multimodal integration studies.[23,]

[24] Therefore, DuAL-Net provides a foundational framework for integrating multiple external modalities into a single WGS-based prediction model, and it can be readily extended to other -omics data and larger genomic regions.

Our ROC analyses across different subset sizes (100, 500, and 1000 SNPs) consistently showed that top-ranked SNPs carried substantially more predictive power than bottom-ranked or randomly selected variants. The top 100 SNPs reached 0.697, while bottom-ranked and random SNPs performed near chance level. This pattern held across most subset sizes, with a gradual convergence observable in the larger subsets (500, 1000 SNPs), where random selection begins to include predictive SNPs by chance. These findings indicate that the combined score derived from DuAL-Net provides a continuous measure of the importance of individual SNPs for AD prediction. Therefore, DuAL-Net offers a supervised feature selection approach that prioritizes features based on their relevance to the prediction task and can be integrated with other models.

Notably, our model identified rs429358 and rs7412 with the highest ranks among all SNPs, which define the *APOE* ε4 genotype and are well-established risk factors for AD.[25-28] In addition, 23 other SNPs, such as rs157588[29], rs11668327[29, 30], and rs157590[29], have also been previously reported to be associated with AD or memory impairment. The presence of these well-validated variants among the top-ranked SNPs provides biological support for our approach. Variants related to lipid metabolism were also found, including rs769455[31, 32], rs34954997[33, 34], rs72654445,[33, 35] rs12721054,[36-38] and rs72654437. Among the SNPs related to lipid metabolism, rs142042446 has been linked to C-reactive protein (CRP), creatinine levels, and glomerular filtration rates (GFR), while rs1064725 has also been associated with CRP.[35, 39] CRP levels and kidney function, represented by creatinine levels or GFR, are potentially associated with AD.[40-43] These findings suggest that our model is capable of detecting both known and potentially novel AD-related variants. However, interpretation should be made cautiously, as the analysis was restricted to a genomic region surrounding *APOE*. Moreover, the sliding window approach may limit the resolution needed to evaluate the individual importance of each SNP.

Although DuAL-Net propose a novel architecture to integrate the information from local SNP window analysis with global annotation-based analysis to predict AD, this study has limitations. First, the current sample size may limit the performance of the trained model. Larger datasets would significantly enhance statistical power and strengthen the model's robustness in capturing a broader range of risk SNPs. Second, the analysis was limited to a specific region around *APOE* on chromosome 19. The TabNet architecture incorporates an attention mechanism, which leads to a sharp increase in computational resource requirements as the input size increases, posing challenges for scalability. To support future expansion of DuAL-Net's input data, incorporating other feature reduction strategies that use linkage disequilibrium[22] or biological pathway to enhance biological relevance may be a promising direction for further development. Third, relying on SNP data curated from Ensembl, which already highlights disease associations, may introduce biases by focusing on well-known loci and risking overlook of novel or less explored SNPs. Future research should expand beyond annotation data alone by integrating truly multimodal data from multiple dimensions in the DuAL-Net architecture.

In conclusion, DuAL-Net presented a promising hybrid framework for AD prediction by effectively integrating both local SNP window analysis and global annotation-based analysis. By prioritizing top-ranked SNPs, it validated superior predictive performance and highlights biological interpretability. Its framework and web server provide an accessible platform for researchers to apply and extend our methodology. Our framework could serve as a foundation for more comprehensive models that integrate diverse biological data types to improve understanding and prediction of complex diseases like AD.